\documentclass[amsmath,nobibnotes,aps,amssymb,pra, prl,aps,showpacs,superscriptaddress,twocolumn, longbibliography, reprint]{revtex4-2}
\usepackage{amsmath,amsfonts,amssymb,amsthm,graphics,graphicx,epsfig,bbm}
\usepackage[colorlinks=true,citecolor=blue,linkcolor=blue,urlcolor=blue]{hyperref}
\usepackage[usenames]{color}
\usepackage{graphicx}
\usepackage{subfigure}
\usepackage{amsmath}
\usepackage{tikz}
\usepackage{booktabs}
\usetikzlibrary{quantikz}
\usepackage{epsfig}
\usepackage{dcolumn}
\usepackage{bm}
\usepackage{color}
\usepackage{times}
\usepackage{epstopdf}
\usepackage{amssymb}
\usepackage{amstext}
\usepackage{latexsym}
\usepackage{float}
\usepackage{hyperref}
\usepackage{amsfonts}
\usepackage{psfrag}
\usepackage{soul,xcolor}
\usepackage[normalem]{ulem}
\usepackage{dsfont}
\usepackage{txfonts}
\usepackage{physics}
\usepackage{footnote}
\usepackage{multirow}
\usepackage{appendix}
\usepackage{mathtools}
\usepackage{algorithm}
\usepackage{algpseudocode}
\usepackage{xspace}     
 
\newtheorem*{theorem*}{Theorem} 
\newtheorem*{lemma*}{lemma}
\newtheorem*{corollary*}{Corollary}
\newtheorem*{remark*}{Remark}

\usepackage{mathtools}

\def\endproof{\hfill$\blacksquare$}

\begin{document}
\setstcolor{red}
 
\title{Non-Kolmogorov-Arnold-Moser Quantum Sensors for Quantum Parameter Estimation} 
\date{\today}

\author{Naga Dileep Varikuti}
\altaffiliation{These authors contributed equally to this work.}			
\affiliation{Pitaevskii BEC Center, CNR-INO and Dipartimento di Fisica, Universit\`a di Trento, Via Sommarive 14, Trento, I-38123, Italy}	
\affiliation{INFN-TIFPA, Trento Institute for Fundamental Physics and Applications, Via Sommarive 14, Trento, I-38123, Italy}
\affiliation{Department of Physics, Indian Institute of Technology Madras, Chennai, India, 600036}
\affiliation{Center for Quantum Information, Computation and Communication, Indian Institute of Technology Madras, Chennai, India 600036}
\author{Sourav Manna}
\altaffiliation{These authors contributed equally to this work.}
\affiliation{Department of Physics, Indian Institute of Technology Madras, Chennai, India, 600036}
\affiliation{Center for Quantum Information, Computation and Communication, Indian Institute of Technology Madras, Chennai, India 600036}
\author{Athreya Shankar}
\affiliation{Department of Physics, Indian Institute of Technology Madras, Chennai, India, 600036}
\affiliation{Center for Quantum Information, Computation and Communication, Indian Institute of Technology Madras, Chennai, India 600036}
\author{Arul Lakshminarayan}
\affiliation{Department of Physics, Indian Institute of Technology Madras, Chennai, India, 600036}
\affiliation{Center for Quantum Information, Computation and Communication, Indian Institute of Technology Madras, Chennai, India 600036}
\author{Vaibhav Madhok}
\email{madhok@physics.iitm.ac.in}
\affiliation{Department of Physics, Indian Institute of Technology Madras, Chennai, India, 600036}
\affiliation{Center for Quantum Information, Computation and Communication, Indian Institute of Technology Madras, Chennai, India 600036}

\begin{abstract}
Non-KAM (Kolmogorov-Arnold-Moser) systems, when subjected to weak time-dependent perturbations, exhibit an abrupt transition to classical chaos through the breakdown of invariant phase-space tori. We showcase the utilization of non-KAM systems in the quantum regime as quantum sensors, leveraging their sensitivity at \textit{resonances}.
Quantum Fisher information (QFI) is a central quantity in quantum parameter estimation theory that measures how much information a quantum state contains about an unknown parameter that is encoded into it. In other words, it quantifies the sensitivity of a quantum state to small changes in that parameter. In this work, through numerical analysis in conjunction with analytical results, we study the performance of the non-KAM systems for quantum sensing applications by computing the QFI. 
We find that the growth of the QFI is remarkably enhanced when the resonance condition is satisfied. For frequency estimation
under Floquet unitary encodings, we derive a transport bound: if the mean
excitation number grows as $\langle\hat n(t)\rangle\sim t^{\alpha}$, the QFI
obeys $I(t)\lesssim t^{2\alpha+2}$.  The quantum kicked harmonic
oscillator, a paradigmatic non-KAM system, realizes the full hierarchy:
localized dynamics ($\alpha=0$) yield quadratic growth, delocalized
diffusion along stochastic webs ($\alpha=1$) yields quartic growth, and
translationally invariant resonances ($\alpha=2$) saturate the bound with
anomalous hexic growth, $I(t)\sim t^{6}$, established analytically at
resonance $R=2$ and numerically at $R=4$. The enhancement stems from
resonance-induced translational symmetry rather than exponential
instability, identifying non-KAM resonances as a metrological resource
distinct from chaos-assisted and criticality-based sensing.
\end{abstract}

\maketitle

\def\endproof{\hfill$\blacksquare$}
\section{introduction}
\label{section-1}

Quantum sensing is one of the most promising applications of near-term quantum technologies~\cite{rosi2014precision, degen2017quantum, RevModPhys.90.035005, DeMille2024QuantumSensing}. By harnessing quantum effects such as entanglement~\cite{giovannetti2004quantum, giovannetti2006quantum}, squeezing~\cite{PhysRevD.23.1693}, and quantum coherence~\cite{RevModPhys.89.041003}, quantum metrology is well poised to probe physical parameters with precision beyond what is achievable with classical approaches. In a typical sensing protocol, an unknown quantity, e.g., a magnetic or electric field strength or a frequency, is encoded in the evolution of a quantum state and subsequently inferred through appropriate measurements and post-processing~\cite{degen2017quantum}. Frequency estimation, in particular, is central to precision measurements in platforms like trapped ions, which offer highly controllable internal and motional degrees of freedom for quantum sensing~\cite{PhysRevLett.110.163604, Wolf2019, PRXQuantum.5.020314, PhysRevX.14.031030}. For a parameter to be estimated precisely, it is desirable for the state of the system to exhibit a pronounced response to small variations in that parameter~\cite{braunstein1994statistical}. This naturally raises the question of whether suitably engineered dynamics can amplify such a response. In the classical regime, non-Kolmogorov-Arnold-Moser (non-KAM) systems provide a natural setting to explore this possibility, owing to their striking sensitivity to weak perturbations~\cite{sankaranarayanan2001chaos, sankaranarayanan2001quantum, varikuti2023probing, varikuti2026instabilities, sreeram2025information, varikuti2024quantum}.

Non-KAM systems are dynamical systems that fall outside the applicability regime of the Kolmogorov-Arnold-Moser (KAM) theorem~\cite{arnold2009proof, kolmogorov1954conservation, moser1967convergent, moser1962invariant}. An \(n\)-degree-of-freedom integrable Hamiltonian system \(H_0\) can be described by generalized action-angle coordinates, with trajectories confined to invariant \(n\)-tori in phase space. When \(H_0\) is non-degenerate, KAM theory establishes that a sufficiently weak and generic perturbation \(\epsilon H_1\) preserves a large set of these invariant tori~\cite{arnold2009proof, kolmogorov1954conservation, moser1962invariant, moser1967convergent, poschel2009lecture, Jesus2026arxive, Kumari2018pre}. The resulting dynamics, therefore, remain predominantly regular despite the perturbation. However, when the conditions required by KAM theory are violated, the persistence of invariant tori is no longer guaranteed, and even arbitrarily weak perturbations can induce widespread destruction of invariant tori and the onset of chaos~\cite{sankaranarayanan2001quantum, sankaranarayanan2001chaos, varikuti2023probing, varikuti2026instabilities, Jesus2026arxive, chinni2022trotter}. Such behavior is characteristic of non-KAM systems. For time-periodic perturbations, near resonances, where the characteristic frequencies of the unperturbed and perturbing dynamics become commensurate, the dynamics display a particularly strong response to variations in system parameters. In the quantum regime, these non-KAM effects can manifest as sharp changes in eigenvalue and eigenvector statistics, as well as in the associated operator dynamics~\cite{varikuti2023probing, varikuti2026instabilities}. 

Motivated by this striking behavior of non-KAM systems, we ask in this work: \textit{Can a non-KAM system be harnessed for quantum sensing purposes by exploiting its sensitivity?} We address this question through rigorous numerical and analytical analysis of the quantum Fisher information (QFI) in the kicked harmonic oscillator, a prototypical non-KAM system. The QFI quantifies the distinguishability of quantum states corresponding to different values of an unknown parameter and sets the ultimate precision permitted by quantum mechanics for its estimation~\cite{braunstein1994statistical}. In this work, we study how the QFI varies with the parameter of interest under resonant and non-resonant conditions, and whether the enhanced dynamical response near resonances translates into improved quantum sensing performance compared to the non-resonant case. Our findings highlight that dynamical instabilities rooted in classical non-KAM dynamics can themselves serve as a resource for quantum metrology, complementing conventional quantum resources such as entanglement, squeezing, and quantum coherence.

This paper is structured as follows. In Sec.~\ref{section-2}, we provide the necessary background for this work. Section~\ref{section-2A} introduces the basics of quantum Fisher information (QFI), while Sec.~\ref{section-2B} introduces the kicked harmonic oscillator (KHO) model and reviews its basic properties. Section~\ref{section-3} presents the main results of this work. In Sec.~\ref{section-3A}, we examine the behavior of the Loschmidt echo and its implications for the QFI, while Sec.~\ref{section-3B} examines how the growth of the mean energy bounds the growth of the QFI. In Sec.~\ref{section-3C}, we study the behavior of the QFI in both resonant and non-resonant regimes. In Sec.~\ref{section-4}, we elucidate the role of translation invariance in the quantum KHO in the growth of the QFI. We then evaluate the classical Fisher information (CFI) associated with quadrature measurements and assess their ability to approach the QFI in Sec.~\ref{section-5}. Finally, in Sec.~\ref{section-6}, we summarize and discuss our findings and outline possible directions for future work.

\section{Background}
\label{section-2}

Here, we briefly introduce the main quantity of interest, the quantum Fisher information, and provide details of the system under study, the kicked harmonic oscillator.  

\subsection{Quantum Fisher information}
\label{section-2A}

Quantum sensing aims to estimate an unknown parameter $\omega$ \cite{helstrom1969quantum}. A typical sensing protocol consists of three basic stages: (i) preparing a probe state $|\psi_{\omega}\rangle$, (ii) measuring the state in a suitable basis, and (iii) extracting an estimate of $\omega$ from the measurement outcomes using an appropriate statistical inference procedure. The precision of an unbiased estimator is bounded by the Cram\'er--Rao inequality \cite{rao1992information, cramer1999mathematical},
\begin{equation}
    \Delta^2\omega \geq \frac{1}{n I(\omega)},
\end{equation}
where $n$ is the number of experimental repetitions and $I(\omega)$ denotes the classical Fisher information (CFI). For a measurement outcome $X$ described by the probability distribution $P_{\omega}(X)$, the CFI is defined as
\begin{equation}
    I(\omega)
    = \mathbb{E}_{\omega}
    \left[
        \left(
            \partial_{\omega}\ln P_{\omega}(X)
        \right)^2
    \right].
\end{equation}
Because the probability distribution depends on the measurement being performed, the CFI is inherently measurement-dependent. Optimizing it over all positive operator-valued measurements (POVMs) yields the quantum Fisher information (QFI), which sets the ultimate precision allowed by the quantum state \cite{braunstein1994statistical, braunstein1996generalized}. 

Geometrically, the CFI corresponds to the unique monotone Riemannian metric on a smooth statistical manifold \cite{chentsov1982statiscal, morozova1991markov}. This implies that for a probability distribution $P_\omega$ parametrized by $\omega$, the second derivative of any valid distance metric $d(P_\omega, P_{\omega+\epsilon})$ with respect to $\epsilon$ becomes equal to the CFI up to a multiplicative factor, i.e., 
\begin{equation}
\partial^2_{\epsilon}d(P_\omega, P_{\omega+\epsilon})|_{\epsilon\rightarrow 0}\equiv C I(\omega), 
\end{equation}
where $C$ is a real number and depends on the distance metric considered. In other words, the Fisher information quantifies the sensitivity of a parametric probability distribution to infinitesimal changes in its parameters \cite{meyer2021fisher}. Similarly, given $|\psi_{\omega}\rangle=\hat{U}_{\omega}|\psi\rangle$, the quantum Fisher information can be identified with the sensitivity of $|\psi_{\omega}\rangle$ to tiny variations in the parameter $\omega$ \cite{helstrom1969quantum, braunstein1994statistical}:
\begin{equation}\label{QFI_ofig}
I(\omega)= 4\lim_{\epsilon\rightarrow 0}\left( \dfrac{1-|\langle\psi_{\omega+\epsilon}|\psi_\omega\rangle|^2}{\epsilon^2} \right)
=\left. -2\partial^2_{\epsilon}|\langle \psi_{\omega+\epsilon}|\psi_\omega \rangle|^2\right|_{\epsilon=0}. 
\end{equation}
In the above equation, the fidelity metric $|\langle\psi_{\omega+\epsilon}|\psi_\omega\rangle|^2$ is assumed to be smooth and differentiable, as is generally the case. Note that for pure states, the fidelity metric is described by the Fubini--Study metric, while for mixed states, it is characterized by the Bures metric. The unknown parameter $\omega$ may characterize various properties of the Hamiltonian, including its characteristic frequency, coupling strengths, and magnetic or electric field strengths. Note that for pure states, the fidelity metric is given by the Fubini-Study metric, whereas for mixed states, the Bures metric is used. Beyond its role in quantum sensor characterization, the QFI serves as an important tool for probing multipartite entanglement in many-body quantum systems~\cite{PhysRevA.85.022321, PhysRevA.85.022322, doi:10.1126/science.1250147, hauke2016measuring}.

The performance of a quantum sensor can be enhanced by invoking quantum resources. Let $m$ denote the total available sensing resources, quantified, for instance, by the number of probes. When the probes are employed independently, the Fisher information typically scales linearly with $m$, $I(\omega)\sim m$, corresponding to the standard quantum limit. However, incorporating genuine quantum effects, such as entanglement during the state preparation in step (i) of the sensing protocol, can result in Heisenberg-limited sensing, $I(\omega)\sim m^2$ \cite{giovannetti2004quantum, giovannetti2006quantum}. The divergent susceptibilities associated with ground states near quantum critical transitions have subsequently motivated their use as a resource for quantum metrology~\cite{zanardi2008quantum, invernizzi2008optimal, ivanov2013adiabatic, tsang2013quantum, macieszczak2016dynamical, rams2018limits, frerot2018quantum, chu2021dynamic, garbe2020critical, garbe2022critical, gietka2022understanding}. More recently, quantum chaotic dynamics have also been explored as a resource for quantum metrology \cite{fiderer2018quantum, liu2021quantum}. Motivated by these developments, we consider non-KAM system dynamics in the quantum regime as a potential resource for enhancing quantum sensing applications.

\subsection{Model: Kicked harmonic oscillator}
\label{section-2B}
Kicked harmonic oscillator (KHO) is a paradigmatic non-KAM system that describes the classical motion of charged particles in a magnetic field with the following Hamiltonian~\cite{chernikov1989symmetry, afanasiev1990width, reichl2021transition, rechester1980calculation, ichikawa1987stochastic, ishizaki1991anomalous, daly1994classical, borgonovi1995translational, zaslavsky2007physics}: 
\begin{eqnarray}
H=H_0+ K\cos(kX)\sum_{n=-\infty}^{\infty}\delta(t-n\tau),
\end{eqnarray}
where $H_0$ denotes the Hamiltonian of a harmonic oscillator with the natural frequency $\omega$, and $K$ denotes the kicking strength. In this work, we fix the wave vector at $k=1$. In action-angle coordinates as given by the canonical transformation $(P, X)=(\sqrt{2J\omega}\sin\theta, \sqrt{2J/\omega}\cos\theta)$, the harmonic oscillator remains classically degenerate meaning that $H_0=\omega J$ is linear in the action variable $J$ while $\theta$ remains a cyclic coordinate. Therefore, $H_0$ is a non-KAM system \cite{poschel2009lecture}. Consequently, the time-periodic perturbations may lead to resonances, which typically occur whenever the natural and the kicking frequencies commensurate, i.e.,  $\omega\tau=2\pi/R$, where $R\in\mathbb{Z}^{+}$.

\textit{Classical dynamics: }When the non-KAM system attains a resonance, its phase-space structure can change dramatically, even if the perturbation is made arbitrarily small. In the KHO model, this transition can be seen clearly from the following two-dimensional map:
\begin{eqnarray}\label{dynmap}
u_{n+1} &=&(u_n+\epsilon\sin v_n)\cos(\omega\tau) +v_n \sin(\omega\tau),\nonumber\\
v_{n+1} &=&-(u_n+\epsilon\sin v_n)\sin(\omega\tau) +v_n \cos(\omega\tau),
\end{eqnarray}
where $u=P/\omega$, $v=X$, and $\epsilon=K/\omega$. 
Note that $H_0$ possesses rotational symmetry in phase space, while $\cos X$ remains invariant under translations of $X$ by integer multiples of $2\pi$. The dynamics described by Eq. (\ref{dynmap}) therefore shows a natural competition between these two symmetries, which becomes particularly evident at the resonant values $R\in R_c\equiv{1,2,3,4,6}$ \cite{chernikov1989symmetry,afanasiev1990width}. For $R=1$ and $R=2$, the map can be solved trivially. For the remaining resonances, the phase space develops stochastic webs whose thickness is exponentially small in $\epsilon$ \cite{afanasiev1990width}. These webs originate near the outer boundaries of the periodic islands and are formed predominantly by their separatrices. Depending on the value of $R$, the resulting phase-space structure exhibits either square or Kagome-like tessellations. Although the KHO is a two-dimensional system, the stochastic webs play a role analogous to Arnold diffusion in systems with more than two degrees of freedom, allowing phase-space trajectories to undergo unbounded diffusion. In particular, trajectories initiated near the separatrices can progressively spread throughout the stochastic web, leading to diffusive growth of the mean energy, $\langle E(t)\rangle\sim\epsilon^2t$, where $t$ denotes the number of time steps.

\begin{figure}
\includegraphics[scale=0.35]{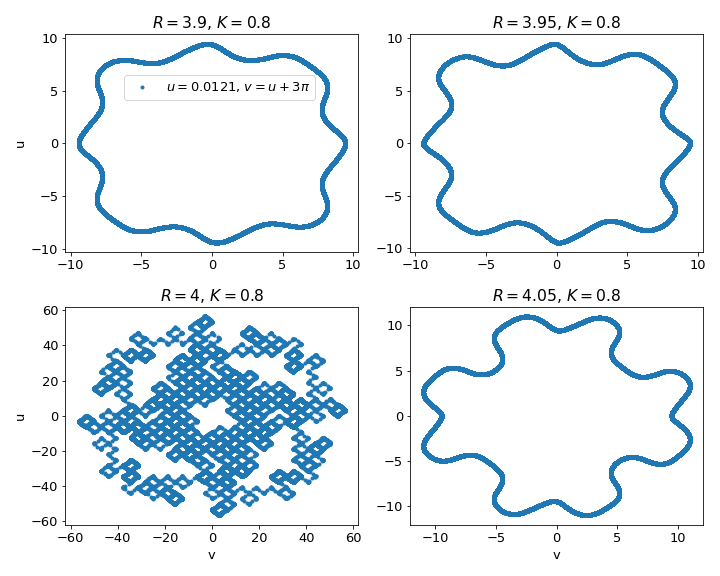}
\caption{\label{fig:poincare} Classical phase-space trajectories for a fixed initial condition at four different values of $R$ in the vicinity of $R=4$. The initial condition is $u=0.0121$ and $v=u+3\pi$. The kicking strength is fixed at $K=0.8$ in all panels. Panels (a), (b), and (d) correspond to the non-resonant regime, where the trajectory associated with the chosen initial condition undergoes only a slight distortion. In contrast, panel (c) corresponds to the resonant case, $R=4$, where the trajectory explores an extended region of phase space and forms a thick stochastic web. }
\end{figure}

In contrast, the non-resonant $R$ will comprise all real numbers except the integers. In this case, for small $\epsilon$, the circular phase space trajectories of the harmonic oscillator are only slightly distorted. Resultantly, the phase space diffusion is suppressed, leading to stagnated energy growth. Figure \ref{fig:poincare} shows the phase space trajectories of the KHO system for a few randomly chosen initial conditions in the vicinity of $R=4$. The corresponding separatrix equation is given by $v=\pm(u+\pi)+2l\pi$, $l\in\mathbb{Z}$ \cite{afanasiev1990width}. The phase space is regular when the system is non-resonant. However, it can be seen from the figure that the trajectories get increasingly deformed as $R$ approaches $4$. Such behavior has applications in the chaotic electron transport in semiconductor superlattices \cite{fromhold2001effects, fromhold2004chaotic}.
 
\textit{Quantum dynamics: }In the quantum regime, the periodicity of the perturbation gives rise to a Floquet description of the KHO dynamics \cite{berman1991problem,shepelyansky1992quantum,daly1994classical,kells2005quantum,daly1996non,engel2007quantum,kells2004dynamical},
\begin{eqnarray}
\hat{U}=\exp\left\{-iK\cos\hat{X}\right\}\exp\left\{-i\omega\tau\hat{a}^{\dagger}\hat{a}\right\},
\end{eqnarray}
where $\hat{X}=(\hat{a}+\hat{a}^{\dagger})/\sqrt{2\omega}$. Throughout this work, we set $\hbar=1$. Ion traps and Bose--Einstein condensates have been proposed as experimentally accessible platforms for realizing the dynamics of the quantum KHO \cite{Gardiner1997,carvalho2004web,gardiner2000nonlinear,duffy2004nonlinear,billam2009quantum}. At the resonant values $R\in R_c$, the translational symmetry of the phase space translates into the existence of one or two families of displacement operators that commute with the $R$-step Floquet operator, $\hat{U}^{R}$ \cite{borgonovi1995translational}. This translational invariance has important consequences for the Floquet spectrum: the corresponding Floquet states are generally extended across phase space, resulting in an unbounded growth of the mean energy, $\langle\hat{a}^{\dagger}\hat{a}(t)\rangle$.


However, even a small deviation of $R$ from the resonant values in $R_c$ can break the translational invariance, causing the Floquet states to become strongly localized. This localization suppresses the quantum dynamics and is also reflected in the behavior of operator scrambling \cite{varikuti2023probing}. Furthermore, in the regime $K\lesssim\pi$, the quantum KHO can be mapped onto an effective tight-binding model, providing an explanation for the dynamical localization observed at irrational values of $R$ \cite{frasca1997quantum,kells2005quantum}. Thus, the quantum KHO exhibits a pronounced sensitivity to variations in $R$ near resonance, particularly in the presence of translational invariance.

Motivated by this sensitivity, we investigate the quantum KHO as a platform for quantum sensing. Specifically, we explore how its dynamics can be exploited for parameter estimation by using the quantum Fisher information (QFI) as a figure of merit. In the next section, we demonstrate the applicability of quantum KHO dynamics to sensing protocols and characterize its performance through the QFI.

\begin{section}{Results}
\label{section-3}
In this work, we are particularly interested in studying the QFI by taking $\omega$, the natural frequency of a particle trapped in a harmonic potential, as the parameter that needs to be estimated.  Throughout, the QFI is always defined with respect to $\omega$; since we work at fixed $\tau$, we often label the dynamics and display the QFI as a function of the commensurability parameter $R=2\pi/\omega\tau$ for convenience. 
For this purpose, we employ the dynamics of the quantum KHO to imprint $\omega$ onto an arbitrary quantum state, i.e., $|\psi_\omega (t)\rangle=\hat{U}^{t}_{\omega}|\psi\rangle$. Under this setting, the evolution of the QFI follows 
\begin{eqnarray}\label{quench}
I(\omega ;|\psi\rangle, t)&=& 4\lim_{\epsilon\rightarrow 0}\left( \dfrac{1-|\langle\psi|\hat{U}_{\omega+\epsilon}^{\dagger t}\hat{U}^{t}_{\omega}|\psi\rangle|^2}{\epsilon^2} \right)\nonumber\\
&=& 4\Delta^2_{|\psi\rangle} \hat{h}^{(t)}_{\omega},
\end{eqnarray}
where $\hat{h}^{(t)}_{\omega} =i \hat{U}^{\dagger t}_{\omega}\partial_{\omega}\hat{U}^{t}_{\omega}$, and $\Delta^2$ denotes variance with respect to the initial state $|\psi\rangle$. Given $|\psi_{\omega}(t)\rangle$, the operator $\hat{h}^{(t)}_{\omega}$ generates infinitesimal translations in the space of $\omega$. As detailed in Appendix~\ref{appendix:a}, for $t=1$, the generator takes the following form:
\begin{eqnarray}\label{generator_exp}
\hat{h}^{\left(1\right)}=i\hat{U}_{\omega}^\dagger \partial_{\omega}\hat{U}_{\omega}
=\tau\hat{n}-\dfrac{K}{2\omega}\hat{A}\sin \hat{A},
\end{eqnarray}
where $\hat{n}=\hat{a}^{\dagger}\hat{a}$ denotes the mean energy operator and $\hat{A}=(\hat{a}e^{-i\omega\tau}+\hat{a}^\dagger e^{i\omega\tau})/\sqrt{2\omega}$. 
Then, the generator corresponding to the $t$-th time step can be written as
\begin{eqnarray}\label{timegen}
\hat{h}^{(t)}_{\omega}&=&\sum_{j=0}^{t-1}\hat{U}^{\dagger j}_{\omega}\hat{h}^{(1)}_{\omega}\hat{U}^{j}_{\omega}\nonumber\\ 
&=&\sum_{j=0}^{t-1}\left[\tau\hat{n}(j)-\dfrac{K}{2\omega}\hat{A}(j)\sin \hat{A}(j)\right],
\end{eqnarray}
where $\hat{n}(j)$ and $\hat{A}(j)$ denote the Heisenberg-evolved forms of the respective operators under the quantum KHO dynamics for $j$-time steps. The time evolution of the QFI is then given by 
\begin{eqnarray}\label{gen-time}
I\left(t\right)=4\;\Delta^{2}_{|\psi\rangle}\sum_{j=0}^{t-1}\left[\tau\left(\hat{a}^\dagger \hat{a}\right)(j)-\dfrac{K}{2\omega}\hat{A}(j)\sin \hat{A}(j)\right]
\end{eqnarray}
In this work, we primarily employ the above equation to numerically compute the QFI for the quantum KHO model.

Note that Eq.~(\ref{gen-time}) suggests a strong connection between the QFI and fluctuations in the mean energy, while Eq.~(\ref{QFI_ofig}) indicates that the QFI is also intimately related to the Loschmidt echo. In the following subsections, we examine these connections in greater detail. In what follows, we first examine the implications of the Loschmidt echo and mean-energy growth for the QFI in the quantum KHO model, and then turn to a numerical analysis of the QFI.

\subsection{Loschmidt Echo}
\label{section-3A}

\begin{figure}
\includegraphics[scale=0.44]{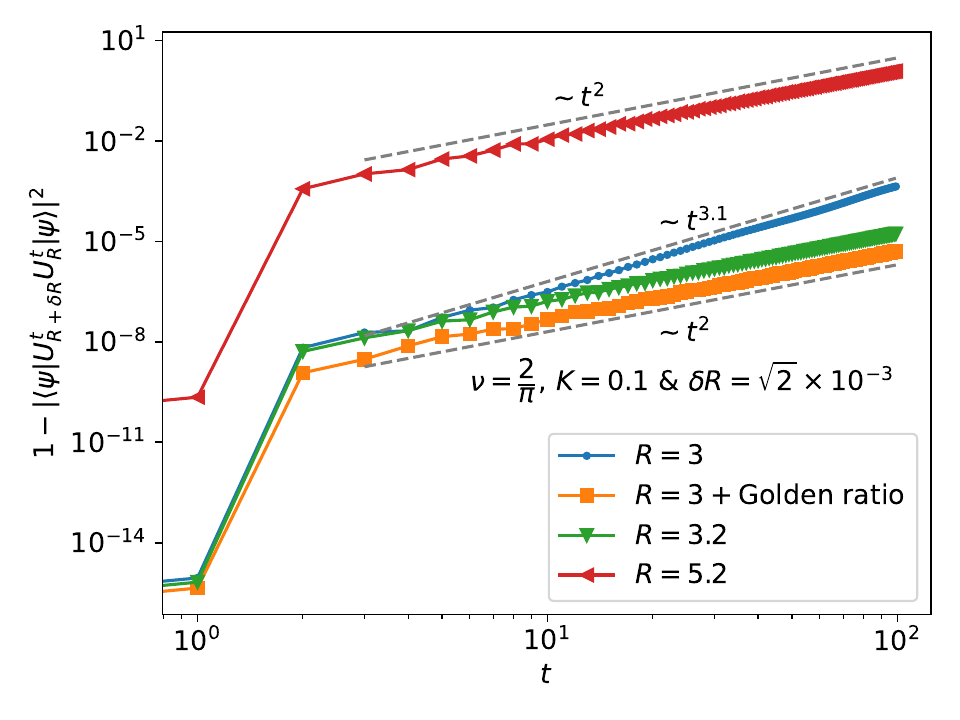}
\caption{\label{fig:Loschmidt} Evolution of the Loschmidt echo for the quantum KHO system near the resonance point. The system is initialized in the vacuum quantum state $|0\rangle$. $R=3$ is the resonance point, and the remaining three cases correspond to dynamical localization. The dashed style lines are plotted to demonstrate the time scalings of the LE.}
\end{figure}

The quantity appearing in the first equality of Eq.~(\ref{QFI_ofig}),
$|\langle\psi|\hat{U}^{\dagger t}_{\omega+\epsilon}\hat{U}^{t}_{\omega}|\psi\rangle|^2$, is the Loschmidt echo (LE), a widely used measure for characterizing the sensitivity of quantum dynamics to perturbations. The LE was introduced by Peres to quantify the stability of quantum evolution under small changes in the Hamiltonian or the corresponding time-evolution operator \cite{peres1984stability}. It has since been extensively studied across a broad range of dynamical regimes, from regular to fully chaotic systems \cite{prosen2002stability,Goussev:2012} and is also relevant for error characterization in quantum technologies~\cite{Sahu2026Loschmidt}. For continuous-variable systems in the weak-perturbation regime, the behavior of the LE depends strongly on the underlying classical dynamics. In regular systems, the LE initially exhibits a parabolic decay, $1-\epsilon^2t^2/\hbar^2$, before crossing over to a Gaussian decay at longer times. The initial parabolic regime typically persists up to times of order $O(1/\epsilon)$, over which the Floquet operator can be approximated by its expansion to second order in the perturbation. In contrast, chaotic systems generally exhibit a linear decay at short times, followed by an exponential decay at later times. These decay laws have been established for infinite-dimensional systems, making them particularly relevant to the quantum KHO, whose Hilbert space is likewise infinite-dimensional.

The connection between the LE and QFI becomes especially direct in the limit of an infinitesimal perturbation, $\epsilon\rightarrow0$. This suggests a direct connection between the well-established behavior of the LE and the growth of the QFI in the regular regime of the KHO. However, it is important to note that the standard LE results assume that the perturbed and unperturbed dynamics belong to the same dynamical regime, i.e., that both systems are either regular or chaotic \cite{prosen2002stability}. On the contrary, this assumption does not generally hold in the vicinity of the resonances. In particular, for $R\in R_c$, the KHO exhibits extended dynamics associated with its translational symmetry, whereas an arbitrarily small detuning of $R$ can induce quantum localization~\cite{kells2005quantum}. The resonant points therefore represent a distinct regime in which the conventional LE results cannot be invoked directly and must be considered separately.


We first perform a brief numerical analysis of the LE for the KHO by varying the parameter $\omega$, while fixing $\tau=1$ and $K=0.1$. The results are shown in Fig.~\ref{fig:Loschmidt}, where we plot $1-|\langle\psi|\hat{U}_{\omega+\delta\omega}^{t}\hat{U}_{\omega}^{t}|\psi\rangle|^2$ as a function of the discrete time for three different values of $\omega$. For $R=3+\sqrt{5}$, which is irrational, the system does not exhibit unbounded diffusion \cite{frasca1997quantum}, and the corresponding classical phase space remains regular. Consistent with the behavior expected for regular systems \cite{prosen2002stability}, the quantity $1-$LE shows quadratic growth with time. A similar behavior is observed for $R=3.2$ and $R=5.2$. In contrast, at the resonance point $R=3$, we observe that $1-$LE grows as $\sim t^{3.1}$.
These results provide an initial indication that the QFI at the resonances can exhibit a substantial enhancement in its temporal scaling. As we discuss below, the origin of this enhancement can be traced to the underlying growth of the mean energy \cite{pang2017optimal}.

\subsection{Mean energy growth vs. QFI}
\label{section-3B}

In atom-optical models, the evolution of the QFI is strongly correlated with the mean energy growth in the system \cite{pang2017optimal, giovannetti2006quantum, garbe2022critical}. If the mean energy growth in those systems scales as $\langle\hat{n}(t)\rangle\sim t^\alpha$, where $\alpha\geq 0$, then the maximum attainable scaling of the QFI follows: $I(\omega ;|\psi\rangle, t)\sim t^{2\alpha+2}$. As we shall show in the following,  this holds for the quantum KHO model as well. To see this, we consider the following inequality.
\begin{eqnarray}\label{bound}
I(\omega; |\psi\rangle, t)=4\Delta^2\left(\sum_{j=0}^{t-1}\hat{h}^{(j)}_{\omega}\right)\leq 4\left(\sum_{j=0}^{t-1}\sqrt{\Delta^2\hat{h}^{(j)}_{\omega}}\right)^2.
\end{eqnarray}
We can further bound the standard deviation of the time evolved generator $\hat{h}_{\omega}(j)$ as 
\begin{eqnarray}
\sqrt{\Delta^2 \hat{h}^{(j)}_{\omega}}\leq\sqrt{\tau^2\Delta^2\hat{n}(j)}+\sqrt{\dfrac{K^2}{4\omega^2}\Delta^2 \left[\hat{A}(j)\sin{\hat{A}}(j)\right]},
\end{eqnarray}
which leads to the following inequality:
\begin{equation}\label{fin-ineq}
I(t)\leq 4 \left(\sum_{j=0}^{t-1}\tau\Delta \hat{n}(j)+\dfrac{K}{2\omega}\Delta \left[\hat{A}(j)\sin \hat{A}(j)\right]\right)^2.
\end{equation}
Eq. (\ref{fin-ineq}) suggests a connection between the maximum possible growth of the QFI and the mean energy fluctuations in the system. From the inequality, one can heuristically argue that in an arbitrary quantum state $|\psi\rangle$, if the mean energy grows as $t^\alpha$, where $\alpha\geq 0$, then $\Delta^2 \hat{n}(t)\lesssim t^{2\alpha}$. Accordingly, the QFI will be bounded by a power-law function of $t$ as implied by the inequality in Eq. (\ref{fin-ineq}):
\begin{equation}
I(\omega; t)\lesssim [\sum_{j=0}^{t-1} O(j^\alpha)]^2.    
\end{equation}
From the following inequality
\begin{equation*}
\int_{0}^{t-1}j^{\alpha}dj<\sum_{j=0}^{t-1} j^\alpha <\int_{0}^{t}j^{\alpha}dj\;,    
\end{equation*} 
it can be immediately shown that $\sum_{j=0}^{t-1} j^\alpha\sim t^{\alpha+1}$, where $\alpha\geq 0$ as mentioned earlier and $t\geq 2$. Therefore, we finally obtain $I(\omega; t)\lesssim O(t^{2\alpha +2})$. Also, note that the second term on the right-hand side of the inequality in Eq. (\ref{fin-ineq}) grows at most as $t^{\alpha+2}$. Thus, the maximum growth of the QFI is solely determined by the mean energy growth. This behavior is corroborated by the results presented in Figs.~\ref{fig:fis-2-trans} and \ref{fig:qfi_mean_energyR4}, which we discuss in detail in the subsequent sections.

To determine how the mean energy grows in the quantum KHO, we consider the Heisenberg evolution of the ladder operators under the quantum KHO dynamics: 
\begin{eqnarray}\label{bosonic-evolution}
\hat{a}(t)e^{i\omega\tau t}=\hat{a}+\dfrac{iKe^{i\omega\tau}}{\sqrt{2\omega}}\sum_{j=0}^{t-1}e^{ij\omega\tau}\sin \hat{A}(j).
\end{eqnarray}
As time progresses, the number of terms in the expansion of a(t) grows linearly. Moreover, all the accumulated terms with time are bounded. This indicates that, under KHO dynamics, the quadrature operators can grow linearly at most, i.e., $\langle X\rangle\sim t$ and $\langle P\rangle\sim t$. Using Eq. (\ref{bosonic-evolution}), one could now bound the mean energy growth of the quantum KHO in an arbitrary state. The bound for the Fock states can be obtained as
\begin{eqnarray}\label{energy-bound}
\langle l|\hat{n}(t)|l\rangle\leq l+\dfrac{\sqrt{2l}Kt}{\sqrt{\omega}}+\dfrac{K^2t^2}{2\omega}. 
\end{eqnarray}
This gives the maximum possible growth of the mean energy for the recurrence dynamics given in Eq. (\ref{bosonic-evolution}). It is known that under the resonance condition, in particular, when the system has translation invariance, the mean energy of the quantum KHO can grow quadratically \cite{borgonovi1995translational}. Consequently, long-time growth of the QFI can display hexic scaling with time at maximum.


\subsection{QFI in the quantum KHO model}
\label{section-3C}

\begin{figure}
\includegraphics[scale=0.33]{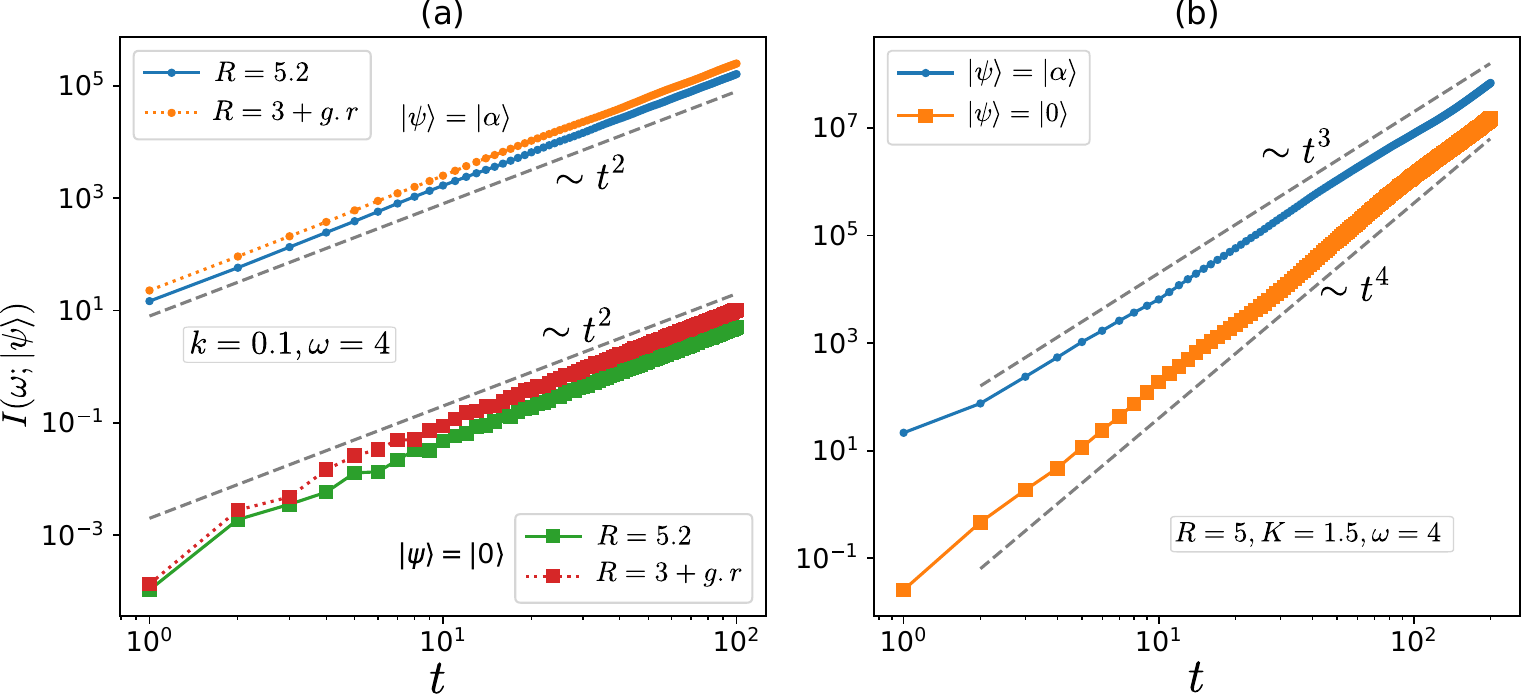}
\caption{\label{fig:fisherfull}Evolution of the QFI for the vacuum quantum state $|0\rangle$ and a coherent state located in phase space at $(x, p)=(-0.5903, -8.1437)$. (a) The QFI is plotted against time at non-resonance points: $R=5.2$ and $R=3+$g.r, where g.r stands for the golden ratio. The growth follows the quadratic law for both states. The dashed lines are plotted to demonstrate the same. (b) The QFI vs time at resonance $R=5$. The same initial states are used as in (a). To ensure delocalization, we take $K=1.5$. We obtain quartic scaling for the vacuum state. The coherent state displays cubic scaling.}
\end{figure}
We first examine the evolution of the QFI for $\omega=2\pi/R$ with non-integer $R$ that are irrational and rational. As mentioned earlier, quantum diffusion is generally suppressed whenever $R$ is non-integer. Hence, the dynamics are completely regular. From known results of the LE \cite{prosen2002stability, Goussev:2012}, the dynamical behavior of the QFI, in this case, could be inferred to be quadratic ($\sim t^2\Delta^2 h_R$). This is indeed confirmed by our numerical results and cross-validated by the LE calculations- see Fig. \ref{fig:Loschmidt}. Moreover, the mean energy, in this case, becomes saturated after some time ($\langle\psi(t)|\hat{a}^\dagger \hat{a}|\psi(t)\rangle\sim c t^0$, $c$ is a constant) \cite{borgonovi1995translational}. Hence, the mean energy viewpoint also supports the above inference that the QFI growth could be at most quadratic. The results are shown in Fig. \ref{fig:fisherfull}(a). The plots show the QFI growth at two non-resonance points ($R=3+g.r$, $5.2$) for two states, namely, the vacuum state $|0\rangle$ and a coherent state $|\alpha\rangle$ with the kicking strength fixed at $K=0.1$. At $K=0.1$, the system is fully regular. 
As Fig.~\ref{fig:fisherfull}(a) illustrates, in this regime, the QFI exhibits quadratic growth with time for both initial states.

We now turn to the resonance points of the quantum KHO. 
Here, we are mainly interested in resonance points $R\notin R_c$, $R\in\mathbb{Z}^+$, that are not translationally invariant in phase space. For this purpose, we fix $R=5$, known to exhibit a quasi-crystalline instance in the classical limit. In this case, the system undergoes a localization-to-delocalization transition as $K$ is tuned away from zero~\cite{shepelyansky1992quantum}. 
As the system becomes delocalized, the mean energy $\langle\psi(t)| \hat{a}^\dagger \hat{a}|\psi(t)\rangle$ grows atmost linearly. 
Following the analysis of the previous subsection, we find that the QFI can grow at most quartically in time, i.e., $I(\omega)\sim t^4$. We verify these predictions numerically for the initial states $|0\rangle$ and $|x+ip\rangle$, where the latter is chosen such that the corresponding phase-space point $(x,p)$ lies on the stochastic web of the classical phase space. The corresponding numerical results are shown in Fig.~\ref{fig:fisherfull}(b). For the vacuum state $|0\rangle$, we observe the predicted quartic scaling of the QFI with time. In contrast, for the coherent state $|x+ip\rangle$, the QFI exhibits a cubic scaling with time. 

In the following, we show that, for $R\in R_c$, the QFI can attain a $\sim t^6$ scaling. This is enabled by the translational invariance present in these cases, which allows the bound given in Eq.~(\ref{energy-bound}) to be saturated. Consequently, the QFI can also exhibit $\sim t^6$ scaling. We discuss this in more detail in the following section.


\end{section}

\begin{section}{Role of translational invariance}
\label{section-4}
The presence of translational invariance implies that $\hat{U}^{R}$ commutes with either a one-parameter or two-parameter group of translations \cite{borgonovi1995translational}. This also leads to extended Floquet states in the phase space and, consequently, to possible unbounded growth of the mean energy. In this case, the mean energy can attain quadratic growth in time, leading to significant enhancements in the growth of QFI. Here we consider $R=1$, $2$, and $4$, where the former two are special cases amenable to exact analytical treatment. The latter case shall be treated both analytically and numerically.

The cases where $R=1$ and $2$ are trivial, for which the Heisenberg evolution of the ladder operators admits a simple form as follows:
\begin{equation}\label{bosonicdyn}
\hat{a}(t)e^{2\pi i t/R}=\hat{a}+i\dfrac{Kt}{\sqrt{2\omega}}\sin\left( \hat{X} \right), \text{ where } R=1\& 2.
\end{equation}
From this, the evaluation of the mean energy growth is straightforward. For instance, in a Fock state $|n\rangle$, the mean energy grows as 
\begin{eqnarray}\label{energy_growth_R=2}
\langle n|(\hat{a}^\dagger \hat{a} )(t)|n\rangle=n+\dfrac{K^2t^2}{4\omega}\left(1-e^{-1/\omega}L_n^0\left(\dfrac{2}{\omega}\right)\right),
\end{eqnarray}
where $L_n^0$ is an associated Laguerre polynomial. As the mean energy growth is quadratic ($\alpha =2$), the QFI will grow algebraically with an exponent $\lesssim 2\alpha+2=6$. Through numerical results and analytical arguments, we indeed confirm that after a sufficiently long time, the evolution of the QFI converges to the scaling $\sim t^6$. 

\begin{figure}
\includegraphics[scale=0.315]{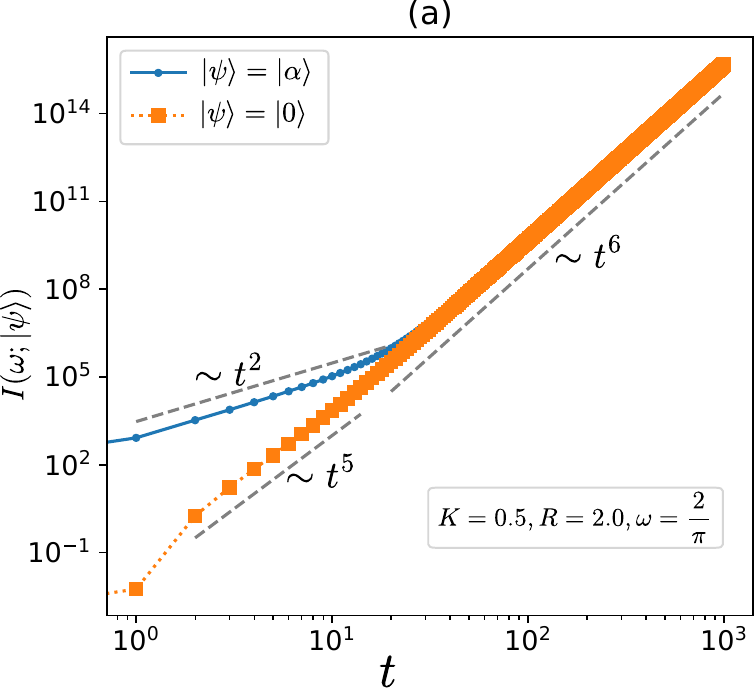}
\includegraphics[scale=0.315]{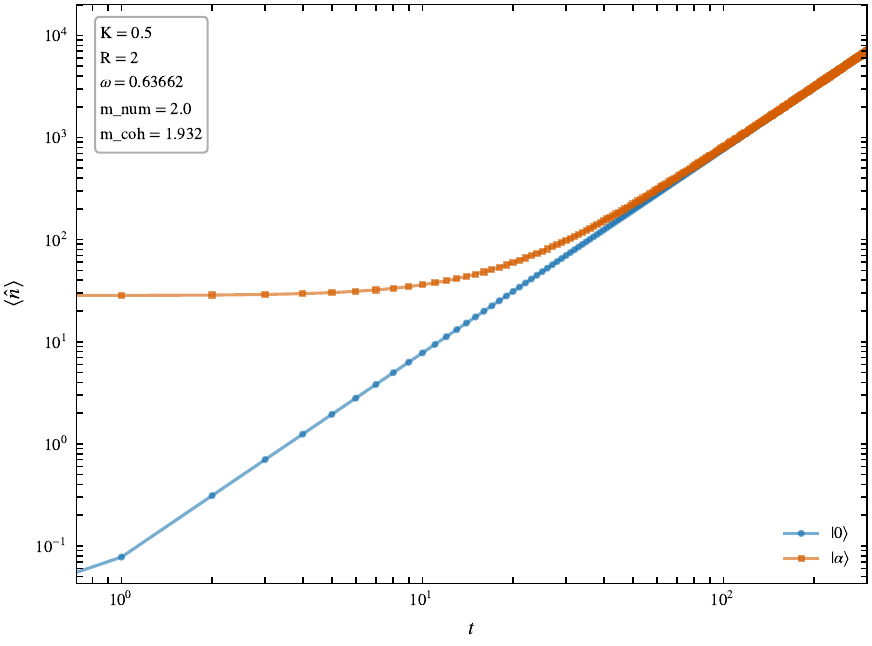}
\caption{\label{fig:fis-2-trans} 
Time evolution of the QFI under translational invariance. (a) QFI at the resonance $R=2$ for the vacuum state $\lvert 0\rangle$ and a coherent state located in phase space at $(x,p)=(0.0121+3\pi,\,0.0121)$. Here $K=0.5$ and $\omega=2/\pi$, with $\tau=\pi^2/2$ so that $\omega\tau=2\pi/R$ at $R=2$. The dashed lines illustrate the respective time scalings of the QFI of the two states. (b) The corresponding growth of the mean excitation number $\langle \hat{n}\rangle$ for the same states, confirming the quadratic energy growth underlying the hexic QFI scaling.}
\end{figure}

\begin{figure}
\includegraphics[scale=0.33]{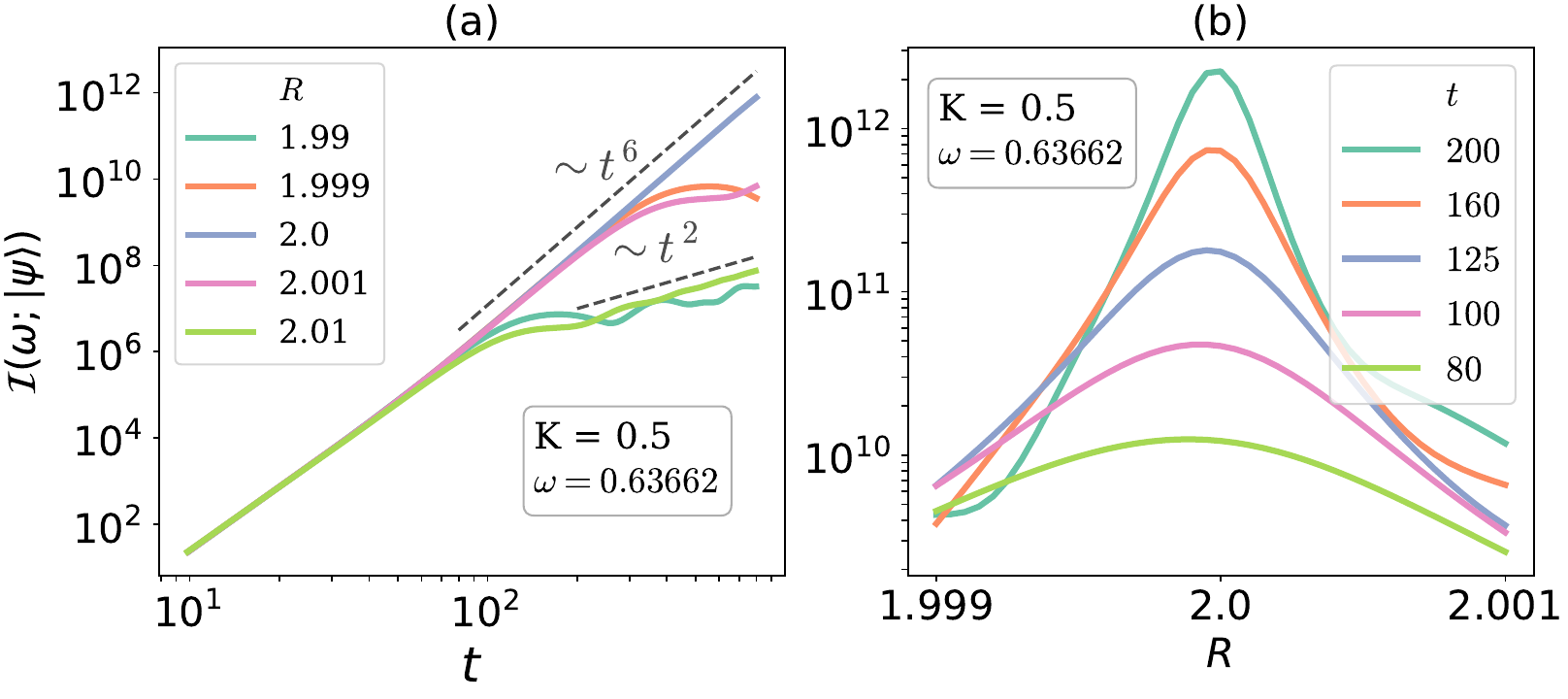}
\caption{\label{fig:qfi_comparison} QFI near the resonance $R=2$, for the initial state $\lvert 0\rangle$, $K=0.5$, and $\omega=2/\pi$, with $\tau = \pi^2/2$ so that $\omega\tau = 2\pi/R$ at $R=2$. (a) QFI as a function of time for different values of $R$ near resonance, illustrating the persistence of the $t^6$ scaling for small detunings over a finite time interval. (b) QFI (with respect to $\omega$) as a function of $R$ at different evolution times, showing the progressive sharpening of the resonance peak with increasing time.}
\end{figure}

To verify the above inference for the QFI, we consider the generator of translations in $\omega$ when $R=2$.
\begin{eqnarray}
\hat{h}_{\omega}=\tau \hat{a}^\dagger \hat{a} -\dfrac{K}{2\omega}\hat{X}\sin\hat{X}.
\end{eqnarray}
The time-evolved generator $\hat{h}_{\omega, t}$ corresponding to $\hat{U}^{t}_{\omega}$ is given in Eq. (\ref{timeevolvedgen}) of Appendix \ref{appendix:b}, wherein the scalar coefficients are polynomials in $t$ with degrees ranging from linear to cubic. This will imply that $I_{|\psi\rangle}(\omega; t)\sim O(t^6)$. We numerically evaluate the QFI for two different initial states, namely, the vacuum state $|0\rangle$ and a coherent state $|\alpha\rangle$. The results are shown in Fig. \ref{fig:fis-2-trans}(a). For the vacuum state, the QFI over a short period exhibits the scaling $\sim t^{5}$. This is then followed by a crossover to $t^6$ scaling at later times. On the other hand, the QFI of the coherent state displays an initial quadratic growth followed by a sharp transition to the Hexic law $t^6$. Although less surprising, the Hexic scaling appears to occur at a similar time scale for both initial states. Moreover, the hexic growth appears after the time $t\sim O(K^{-2/3})$ scaling.

Progress towards a concrete metrological task follows from the observation that the resonance condition is controlled by the product $\omega\tau$. In the present setting, the oscillator frequency $\omega$ is the parameter to be estimated, while the kicking period $\tau$ can, in principle, be externally controlled. Thus, the resonance condition can be approached not only by tuning the oscillator frequency, but also by adjusting the kicking period. For an initial estimate $\omega_0$ of the unknown frequency, one can choose
$\tau \simeq 2\pi/(R\omega_0),
\qquad R\in\mathbb{Z}^{+},$
thereby bringing the system close to a resonant point. Since the QFI is strongly enhanced in the vicinity of the resonance, this adjustment can increase the available information about $\omega$ and consequently improve the precision of a subsequent estimation. This suggests a natural adaptive sensing protocol: a coarse measurement first provides an initial estimate of $\omega$, which is then used to tune $\tau$ closer to a resonance, followed by a second sensing stage with enhanced QFI. Repeating this procedure could progressively concentrate the operation of the sensor around the resonant condition.

The same mechanism suggests a complementary application in frequency stabilization and drift detection. Suppose that the kicking period is fixed such that a known operating frequency $\omega_0$ satisfies
$\omega_0\tau=2\pi/R $.
The system then operates at a translationally invariant resonance, where the QFI exhibits its strongest temporal enhancement. A small change $\omega_0\rightarrow\omega_0+\delta\omega$ moves the system away from the resonant condition. Because the dynamics are highly sensitive to this detuning, the resulting change in the quantum state, and consequently in the measured Fisher information or an appropriate observable, can provide a sensitive indicator of a frequency drift. In this sense, the resonant KHO can be viewed not only as a sensor for estimating an unknown frequency, but also as a dynamical detector for small deviations from a prescribed operating frequency. Such a resonance-based scheme could be useful for monitoring the stability of an oscillator or trapping frequency, where the goal is to detect small changes from a target value rather than continuously estimate the frequency over a broad range.

\begin{figure}
\includegraphics[scale=0.33]{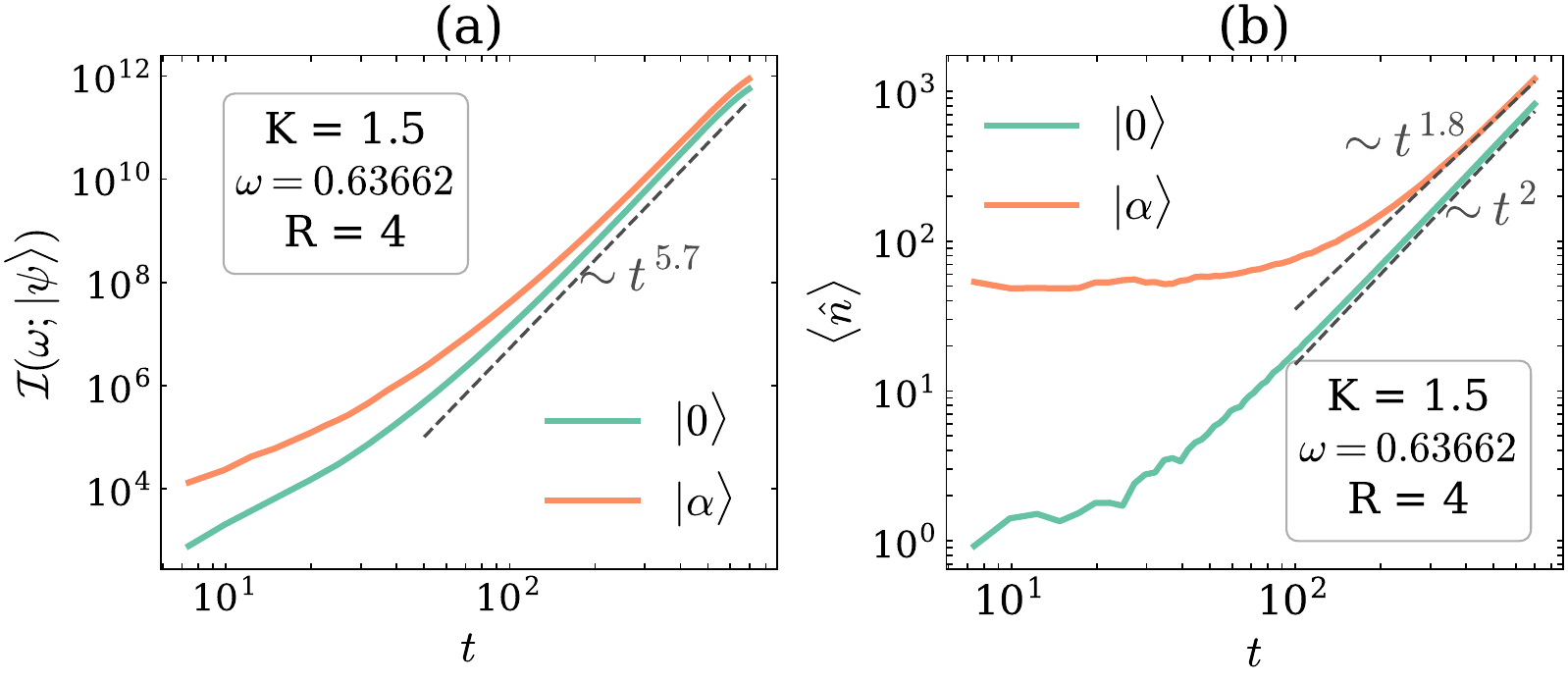}
\caption{\label{fig:qfi_mean_energyR4} 
Time evolution of the QFI at the resonance $R=4$. (a) QFI for the vacuum state $\lvert 0\rangle$ and a coherent state located in phase space at $(x,p)=(-0.5903,\,-8.1437)$. Here $K=1.5$ and $\omega=2/\pi$, with $\tau=\pi^2/4$ so that $\omega\tau=2\pi/R$ at $R=4$. (b) The corresponding growth of the mean energy $\langle \hat{n}\rangle$ for the same states. The dashed lines illustrate the respective time scalings of the QFI and mean energy.}

\end{figure}

An experimentally relevant question is whether the enhanced QFI scaling observed at the resonant values of $R$ requires exact tuning to an integer value. In practice, such precise tuning may be difficult to achieve, and it is therefore important to examine the behavior of the QFI for small detunings from resonance. Figure~\ref{fig:qfi_comparison}(a) shows the time dependence of the QFI for $R$ values in the vicinity of the resonance $R=2$. Interestingly, the resonant scaling is not restricted to the exactly tuned point: for values of $R$ sufficiently close to $2$, the QFI initially follows the same scaling as the resonant case. The agreement persists over an increasingly long time interval as the detuning $|R-2|$ is reduced. For larger detunings, the departure from the resonant scaling occurs at earlier times, indicating that the finite-time dynamics can resolve the deviation from resonance only after a sufficiently long evolution.

This finite-time behavior is further illustrated in Fig.~\ref{fig:qfi_comparison}(b), where the QFI is plotted as a function of $R$ at several evolution times. At each time, the QFI exhibits a pronounced maximum around the resonant value $R=2$, while the peak becomes progressively narrower as the evolution time is increased. Thus, longer evolution allows the dynamics to resolve increasingly small deviations from the resonant condition. Equivalently, two values of $R$ that are indistinguishable within the available evolution time can become distinguishable after sufficiently long evolution because the accumulated dynamical phase carries information about their small difference. These results demonstrate that the enhanced resonant scaling is robust against finite detuning over experimentally relevant time scales, while also highlighting the increasing parameter resolution afforded by longer interrogation times.

We now focus on a more intricate case of $R=4$. The same reasoning can be extended to the other cases where $R=3$ and $R=6$. As discussed in Ref.~\cite{borgonovi1995translational}, the translational symmetries of the Floquet dynamics imply that $\hat{U}_\tau^R$ possesses either a one- or a two-parameter translational invariance, depending on the value of $\omega$. For $R=4$, the annihilation operator involves the sequence of operators $\sin\hat{X}(j)$, with $j=0,\ldots,t-1$ [see Eq.~(\ref{bosonic-evolution})]. We therefore seek the simplest condition for $\omega$ under which the relation $[\sin\hat{X},\hat{U}_\tau^4]=0$ holds. For this purpose, the fourth power of the Floquet operator can be rearranged into the following form:
\begin{equation}
\hat{U}^4
=
\left(
e^{-iK\cos(\hat{P}/\omega)}
e^{-iK\cos\hat{X}}
\right)^2.
\end{equation}
Since $\sin\hat{X}$ is a function of $\hat{X}$, it commutes trivially with the factor involving $\cos\hat{X}$. Hence, the nontrivial condition is that $[\sin\hat{X},\cos(\hat{P}/\omega)]=0$. This can be made to satisfy by taking the discrete set of frequencies $\omega={1}/{(2k\pi)}$, where $k\in\mathbb{Z}^{+}$. The resulting symmetry of the Floquet evolution gives
$
\hat{U}^{\dagger 4}\sin\hat{X}\,\hat{U}^{4}
=
\sin\hat{X}.
$
More explicitly, for $j=0,1,2,3$, $\sin\hat{X}(j)=\sin\hat{X}(t+j)$ for $t=4s,$
where $s$ is a non-negative integer. This is also called the characteristic quantum-resonance condition of the system~\cite{billam2009quantum}. It then follows that for $t=4s$, Eq.~(\ref{bosonic-evolution}) transforms as
\begin{eqnarray}\label{R4bos}
\hat{a}(t)
=
\hat{a}
+
\frac{iKt}{4\sqrt{2\omega}}
\sum_{j=0}^{3}
e^{ij\pi/2}\sin\hat{X}(j).
\end{eqnarray}
As a result, the mean energy operator evolves in the Heisenberg picture as 
\begin{eqnarray}
\left(\hat{a}^{\dagger}\hat{a}\right)(t) = \hat{a}^{\dagger}\hat{a} + \dfrac{iKt}{4\sqrt{2\omega}}\left( \hat{a}^{\dagger}\hat{B}-\hat{B}^{\dagger} \hat{a} \right) + \dfrac{K^2t^2}{32\omega} \hat{B}^{\dagger}\hat{B},  
\end{eqnarray}
where $\hat{B} = \sum_{j=0}^{3}e^{ij\pi/2}\sin\hat{X}(j)$. For a generic state $|\psi\rangle$, the above expression implies that the mean energy grows quadratically with discrete time unless $\langle \psi | \hat{B}^{\dagger}\hat{B} |\psi\rangle = 0$. Consequently, the QFI can grow with at most $\sim t^6$ scaling.

We confirm this behavior through numerical simulations for both the vacuum state and a coherent state. The results are shown in Fig.~\ref{fig:qfi_mean_energyR4}, where the QFI and mean energy are plotted in panels~\ref{fig:qfi_mean_energyR4}(a) and~\ref{fig:qfi_mean_energyR4}(b), respectively. While the mean energy grows approximately quadratically in time, the QFI exhibits a $\sim t^{5.7}$ scaling, in close agreement with our analytical prediction of $t^6$ scaling. These results demonstrate that translational invariance induced by resonance can substantially enhance the sensing capability of the system, without requiring conventional quantum resources such as entanglement or squeezing. This reveals dynamical symmetry as an alternative route toward enhanced quantum sensing.

\end{section}

\begin{section}{Quadrature measurements}
\label{section-5}

\begin{figure}
\includegraphics[scale=0.39]{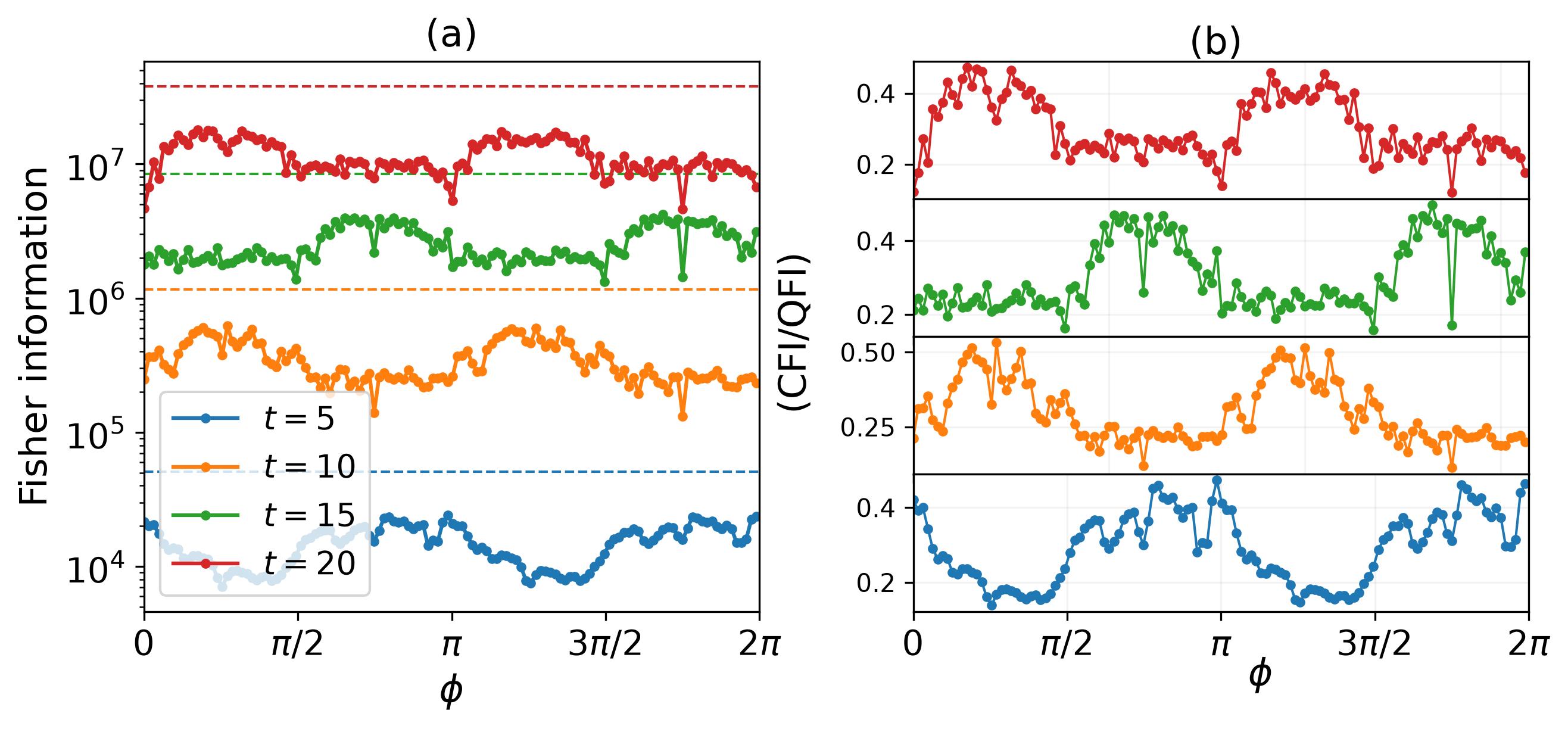}
\caption{ Comparison of the quantum Fisher information (QFI) and classical Fisher information (CFI) as functions of the quadrature angle $\phi$ at the resonance $R=4$, for an initial coherent state with $(x,p)=(\pi,0)$ at discrete time steps $t=5,10,15,$ and $20$. We fix $K=0.1$ and $\omega=1/(2\pi)$, with $\tau$ determined by $\omega\tau=2\pi/R$ for $R=4$. (a) The thick curves with markers $\cdot$ denote the CFI, while the dashed lines denote the QFI. (b) The ratio $\mathrm{CFI}/\mathrm{QFI}$ as a function of $\phi$ for the same time steps. The results show that the CFI remains of the same order of magnitude as the QFI across the quadrature angles considered, indicating that quadrature measurements can attain a substantial fraction of the QFI. }

\label{fig:CFIvsphi}
\end{figure}

While the quantum Fisher information sets the maximum achievable precision for estimating an unknown parameter in the Hamiltonian under an optimal measurement, identifying a measurement strategy that saturates the quantum Cramér–Rao bound is generally challenging. Here, we consider quadrature measurements and examine the classical Fisher information (CFI) associated with estimating $\omega$, as they are a natural choice in atom-optical settings such as the KHO. The quadrature operator is given by $Q_\phi=\left(e^{-i\phi}\hat{a}^\dagger+e^{i\phi}\hat{a} \right)/\sqrt{2}$. By tuning the phase angle $\phi$, we examine how CFI varies at different discrete time. Let $\prod_{q_\phi}$ be the projector along an eigenstate $|q_\phi\rangle$ of the quadrature operator $Q_\phi$. Then the CFI can be computed using the following relation:
\begin{eqnarray}\label{CFI}
I_C(Q_\phi) =\sum_{q}\dfrac{1}{p(q_\phi|\omega)}\left(\dfrac{\partial p(q_\phi|\omega)}{\partial \omega} \right)^2, 
\end{eqnarray}
where $P(q_\phi|\omega)=\langle \psi_\omega|\prod_{q_\phi}|\psi_\omega\rangle=\langle \psi|\hat{U}_{\omega}^{\dagger n}\prod_{q_\phi}\hat{U}^{n}_{\omega}|\psi\rangle$. The summation is performed over all the eigenstates $\{|q_\phi\rangle\}$ of $Q_{\phi}$. Moreover $\partial_{\omega} p(q_\phi|\omega)=2\operatorname{Im}(\langle\psi|\hat{U}_{\omega}^{\dagger n}\prod_{q_\phi}\hat{U}^{n}_{\omega}h_{\omega}(n)|\psi\rangle)$, where $h_\omega(n) = i \hat{U}^{\dagger n}\partial_{\omega}\hat{U}^{n}$ bears the same meaning as before; the generator of translations in parameter space of $\omega$.

To assess the performance of quadrature measurements, we compare the CFI with the QFI as a function of the quadrature angle $\phi$ at the resonance $R=4$. For an initial coherent state with $(x,p)=(\pi,0)$, we evaluate both quantities at several discrete time steps, $t=5,10,15,$ and $20$, as shown in Fig.~\ref{fig:CFIvsphi}. The CFI exhibits an angle dependence while remaining of the same order of magnitude as the QFI across the range of quadrature angles considered. This is further illustrated by the ratio $\mathrm{CFI}/\mathrm{QFI}$, which shows that quadrature measurements can attain a substantial fraction of the QFI. Thus, even though the quadrature measurements do not necessarily realize the optimal measurement, they can capture a significant portion of the available quantum Fisher information.


\end{section}

\begin{section}{Discussion}
\label{section-6}
Non-Kolmogorov-Arnold-Moser systems are known to provide a faster route to classical chaos through an abrupt breaking of the invariance phase space tori. In this work, we have demonstrated that non-KAM systems as promising candidates for the quantum sensing applications by exploiting their extreme sensitivity near resonance conditions.
By analyzing the quantum Fisher information of the kicked harmonic oscillator under unitary parameter encoding, we showed that the breakdown of the KAM assumptions can lead to a substantial enhancement of parameter sensitivity compared to the non-resonant regimes in the parameter space. The observed enhancement originates from the amplified response of the quantum dynamics near resonances, where small perturbations induce significant changes in the evolved state.


The central result of this work is that the temporal growth of the QFI is strongly controlled by the dynamical growth of the mean energy. The accumulation of the generator over successive kicks provides a natural mechanism for enhancing the distinguishability of states corresponding to nearby values of the oscillator frequency. This establishes a direct connection between dynamical energy growth and metrological sensitivity. In particular, if the mean excitation number grows algebraically as $\langle \hat n(t)\rangle\sim t^\alpha$, the QFI is bounded by a scaling of the form $I(t)\lesssim t^{2\alpha+2}$. The enhancement of the QFI is therefore not an independent feature of the quantum dynamics, but is intimately connected to the ability of the system to continuously accumulate energy under the resonant driving.

Three features distinguish non-KAM metrology from its critical and chaotic
counterparts. First, the resource is a \emph{symmetry}: resonance-induced
translational invariance yields ballistic transport and hexic QFI growth,
whereas Lyapunov instability yields only transient gains
\cite{fiderer2018quantum} and generic delocalization only quartic scaling.
Second, the scaling is asymptotic and algebraic, not cut off at an
Ehrenfest time. Third, the protocol is passive --- a fixed kick sequence
--- yet attains the generator growth that otherwise requires optimal
time-dependent control \cite{pang2017optimal}. The KHO has been realized with cold atoms and
proposed for ion traps \cite{Gardiner1997, duffy2004nonlinear, billam2009quantum}, where
$\omega$-estimation at a translationally invariant resonance amounts to
operating the trap at commensurate kick periods.

An important direction for future work is to investigate whether the enhanced QFI scaling observed in the KHO persists in the presence of realistic experimental imperfections, such as decoherence, noise, and fluctuations in the kicking parameters. It would also be interesting to explore whether similar resonance-induced enhancements occur in other non-KAM systems, where the non-KAM behavior arises not from degeneracies in the unperturbed Hamiltonian, but from the breakdown of other KAM assumptions~\cite{sankaranarayanan2001chaos, sankaranarayanan2001quantum}. On the metrological side, extending the analysis to experimentally accessible measurements beyond quadratures and identifying protocols that optimally exploit the enhanced QFI would further help understand the sensing potential of these systems. Finally, since QFI is intimately connected to higher-order correlation functions~\cite{PhysRevLett.120.040402}, which can serve as signatures of quantum resources such as entanglement and coherence~\cite{10.1063/5.0191140, varikuti2025deep}, it would be interesting to ask whether the quantum resources underlying genuine computational advantages can also lead to corresponding metrological advantages.

\end{section}

\begin{acknowledgments}


NDV thanks Philipp Hauke for insightful discussions on Fisher information and quantum resources and for valuable comments on the manuscript. This work was supported by the Provincia Autonoma di Trento and Q@TN, the joint lab between the University of Trento, FBK—Fondazione Bruno Kessler, INFN—National Institute for Nuclear Physics, and CNR—National Research Council. We acknowledge funding by ANRF file number ANRF/ARGM/2025/002679/TS. A.S. acknowledges support by the Department of Science and Technology, Govt. of India through the INSPIRE Faculty Award (DST/INSPIRE/04/2023/001486), by the Anusandhan National Research Foundation (ANRF), Govt. of India through the Prime Minister’s Early Career Research Grant (PMECRG) (ANRF/ECRG/2024/001160/PMS), by IIT Madras through the New Faculty Initiation Grant (NFIG). We acknowledge funding support from the National Quantum Mission, an
initiative of the Department of Science and Technology, Govt. of
India.
We also acknowledge the support provided by the Foundation for QC
Innovation (FQCI), DST-NQM T-Hub at IISc Bengaluru, in facilitating
this project.


\end{acknowledgments}

\bibliography{main.bib}
\onecolumngrid
\appendix

\section{Calculating Generator $\hat{h}^{(t)}_{\omega}$}
\label{appendix:a}
The floquet unitary is $\hat{U} = \exp\left\{{-iK \cos{\hat{X}}}\right\}\exp\left\{{-i\omega\tau \hat{n}}\right\}$, where $\hat{n} = \hat{a}^\dagger a$ and $\hat{X} = (a+a^\dagger)/\sqrt{2\omega}$. Let's say $\hat{U} = \hat{V}(\omega)\cdot\hat{F}(\omega)$, so, $\hat{V}(\omega) = \exp\{{-iK \cos{\hat{X}}}\}$ and $F(\omega) = \exp\{{-i\omega\tau \hat{n}}\}$. The generator for $t=1$ is, $\hat{h}^{(1)}_{\omega} = iU^\dagger \partial_\omega U$.
\begin{equation}
    \label{eq:apx_du}
    \begin{split}
        \partial_\omega \hat{U} & = (\partial_\omega \hat{V})\cdot \hat{F} + \hat{V} (\partial_\omega \cdot \hat{F})\\
        & = \left( \frac{iK}{2\omega}\hat{X}\sin{\hat{X}}\cdot\hat{V}\right)\cdot \hat{F} + \hat{V}\cdot \left( -i\tau \hat{n}\hat{F} \right)
    \end{split}
\end{equation}
\begin{equation}
    \label{eq:apx_hhat}
    \begin{split}
        \hat{h}^{(1)}_{\omega} & = i\hat{F}^\dagger \hat{V}^\dagger \left[ \left( \frac{iK}{2\omega} \hat{X}\sin{\hat{X}}\cdot\hat{V}\right)\cdot \hat{F} + \hat{V}\cdot \left( -i\tau \hat{n}\hat{F} \right) \right]\\
         & = \tau \hat{n} -\frac{K}{2\omega}\hat{F}^\dagger \left( \hat{X}\sin{\hat{X}}\right)\hat{F} \\
        & =  \tau \hat{n} - \frac{K}{2\omega}(\hat{F}^\dagger \hat{X}\hat{F})(\hat{F}^\dagger \sin{\hat{X}}\hat{F})\\
        & = \tau \hat{n} - \frac{K}{2\omega} \left(\frac{e^{-i\omega\tau}\hat{a}+e^{i\omega\tau}\hat{a}^\dagger}{\sqrt{2\omega}}\right)\sin{\left(\frac{e^{-i\omega\tau}\hat{a}+e^{i\omega\tau}\hat{a}^\dagger}{\sqrt{2\omega}}\right)}; \quad \text{using } \hat{F}^\dagger\hat{a}\hat{F} = e^{-i\omega \tau}\hat{a}\\
        \Rightarrow \hat{h}^{(1)}_{\omega} & = \tau \hat{n} - \frac{K}{2\omega} \hat{A}\sin{\hat{A}},
    \end{split}
\end{equation}
where, $\hat{A} = (e^{-i\omega\tau}\hat{a}+e^{i\omega\tau}\hat{a}^\dagger)/\sqrt{2\omega}$. The time evolution of the generator is,

\begin{equation}
    \label{eq:generator_t_def}
    \hat{h}^{(t)}_{\omega} = i (\hat{U}^t)^\dagger \partial_\omega (\hat{U}^t).
\end{equation}
Using the product rule for the derivative of the $t$-step unitary evolution, we have:
\begin{equation}
    \partial_\omega (\hat{U}^t) = \sum_{j=0}^{t-1} \hat{U}^{t-1-j} (\partial_\omega \hat{U}) \hat{U}^j.
\end{equation}
Substituting this into the definition of the generator yields:
\begin{equation}
    \label{eq:generator_t_sum}
    \begin{split}
        \hat{h}^{(t)}_{\omega} &= i (\hat{U}^\dagger)^t \sum_{j=0}^{t-1} \hat{U}^{t-1-j} (\partial_\omega \hat{U}) \hat{U}^j \\
        &= \sum_{j=0}^{t-1} (\hat{U}^\dagger)^{j+1} (i\partial_\omega \hat{U}) \hat{U}^j \\
        &= \sum_{j=0}^{t-1} (\hat{U}^\dagger)^j \left( i \hat{U}^\dagger \partial_\omega \hat{U} \right) \hat{U}^j \\
        &= \sum_{j=0}^{t-1} (\hat{U}^\dagger)^j \hat{h}^{(1)}_{\omega} \hat{U}^j.
    \end{split}
\end{equation}
Using the explicit form of $\hat{h}(\omega; 1)$ derived in Eq.~\eqref{eq:apx_hhat}, we can express the generator at time $t$ in terms of Heisenberg-evolved operators:
\begin{equation}
    \label{eq:generator_t_final}
    \begin{split}
        \hat{h}^{(t)}_{\omega} &= \sum_{j=0}^{t-1} (\hat{U}^\dagger)^j \left[ \tau \hat{n} - \frac{K}{2\omega} \hat{A}\sin{\hat{A}} \right] \hat{U}^j \\
        &= \sum_{j=0}^{t-1} \left[ \tau \hat{n}(j) - \frac{K}{2\omega} \hat{A}(j)\sin{\hat{A}(j)} \right],
    \end{split}
\end{equation}
where $\hat{n}(j) = (\hat{U}^\dagger)^j \hat{n} \hat{U}^j$ and $\hat{A}(j) = (\hat{U}^\dagger)^j \hat{A} \hat{U}^j$. Because $\hat{A}$ is linear in $\hat{a}$ and $\hat{a}^\dagger$, its Heisenberg evolution takes the convenient form:
\begin{equation}
    \hat{A}(j) = \frac{e^{-i\omega\tau}\hat{a}(j) + e^{i\omega\tau}\hat{a}^\dagger(j)}{\sqrt{2\omega}},
\end{equation}
with $\hat{a}(j) = (\hat{U}^\dagger)^j \hat{a} \hat{U}^j$.


\section{Time evolution of the QFI for $R=1$ $\&$ $2$}
\label{appendix:b}
We obtain an analytical expression for the generator for the two special cases of $R$, that are $R=1$ $\&$ $R=2$, as they are amenable to exact computations. For the other cases of unperturbed frequencies, the QFI is obtained via numerical simulations. The local generator for the translations in the parameter space for the time-evolved Floquet operator $U^t$ is given by
\begin{eqnarray}\label{tevolgen}
h_{\omega}(t)=\sum_{j=0}^{t-1}U^{\dagger j}h_{\omega} U^{j},
\end{eqnarray}
where $h_{\omega}$ is the local generator as given in Eq. (\ref{eq:apx_hhat}). We first write down the time evolution of the bosonic ladder operators under the dynamics of the kicked harmonic oscillator when the unperturbed frequency takes the values $\omega=2\pi$ and $\pi$, and $n\pi$ in general for $n\in\mathbb{Z}^+$.
\begin{eqnarray}\label{timeevolvedbosonic}
a(t)e^{i\omega t}=a+i\dfrac{Kt}{\sqrt{2\omega}}\sin\left( \hat{X} \right).
\end{eqnarray}
The time-evolved bosonic ladder operators have linear time dependence for the above-specified frequencies. The result in Eq. (\ref{timeevolvedbosonic}) is also valid for any frequency that is an integer multiple of $\pi$. The linear time dependence implies an unbounded quadratic growth of the energy of the kicked harmonic oscillator at resonances. On the contrary, the growth of the energy operator $\langle a^\dagger a\rangle$ remains linear if the resonance condition is not fulfilled. This can be traced back to the fact that $\langle a\rangle\sim\sqrt{t}$ for non-resonant cases. We now compute the time evolution of the QFI for the special cases of $R=1$ and $2$. The generator becomes
\begin{eqnarray}\label{gen}
\hat{h}_{\omega}&=&\tau \hat{a}^\dagger \hat{a}-\dfrac{K}{2\omega}\left( \dfrac{\hat{a}e^{-i\omega\tau}+\hat{a}^\dagger e^{i\omega\tau}}{\sqrt{2\omega}} \right)\sin\left( \dfrac{\hat{a}e^{-i\omega\tau}+\hat{a}^\dagger e^{i\omega\tau}}{\sqrt{2\omega}} \right)\nonumber\\
&=&\tau \hat{a}^\dagger \hat{a}-\dfrac{K}{2\omega}\hat{X}\sin\hat{X},
\end{eqnarray}
where we used $\omega\tau=2\pi/R$. For some $j\in \mathbb{Z}^+$, 
\begin{eqnarray}
\hat{h}_{\omega}(j)&=&\tau \hat{n}(j)-\dfrac{K}{2\omega}\hat{X}(j)\sin\hat{X}(j)\nonumber\\
&=&\tau\left\{ \hat{a}^\dagger \hat{a} -i\dfrac{Kj}{\sqrt{2\omega}}\sin(\hat{X}) \hat{a}+i\dfrac{Kj}{\sqrt{2\omega}}\hat{a}^\dagger\sin\hat{X} +\dfrac{K^2j^2}{2\omega}\sin^2\hat{X} \right\}-\dfrac{K}{2\omega}\hat{X}\sin\hat{X}
\end{eqnarray}
The time evolution is given by $h_R(t)=\sum_{j=0}^{t-1}U^{\dagger j}h_RU^j$. The time-evolved generator after $t$ time steps becomes
\begin{eqnarray}\label{timeevolvedgen}
\hat{h}_{\omega, t}=\tau\left\{t\hat{a}^\dagger \hat{a} -\dfrac{iKS_1(t)}{\sqrt{2\omega}}\left(\sin (\hat{X}) \hat{a}-\hat{a}^\dagger\sin\hat{X}\right)+\dfrac{K^2S_2(t)}{2\omega}\sin^2 \hat{X} \right\}-\dfrac{Kt}{2\omega}\hat{X}\sin\hat{X} ,
\end{eqnarray}
where 
\begin{eqnarray*}
S_1(t)=\dfrac{t(t-1)}{2}\hspace{0.5cm}\text{and}\hspace{0.5cm}S_2(t)=\dfrac{t(t-1)(2t-1)}{6}. 
\end{eqnarray*}

From Eq. (\ref{timeevolvedgen}), the QFI for any arbitrary number state $|n\rangle$ can be straightforwardly obtained. 
\begin{eqnarray}
I(|n\rangle; R)=4\left( \langle n|h_{R}(t)^2|n\rangle-\langle n|h_R(t)|n\rangle^2 \right) 
\end{eqnarray}
The resulting expression will be a sixth-degree polynomial of $t$, the total discrete time. 

\end{document}